\documentclass[draft]{agujournal2019}
\usepackage{url} %this package should fix any errors with URLs in refs.
\usepackage{lineno}
\usepackage{tikz}
\usetikzlibrary{matrix}
\usepackage{amsmath}
\usepackage[dvipsnames]{xcolor}
\usepackage[inline]{trackchanges} %for better track changes. finalnew option will compile document with changes incorporated.
\usepackage{soul}
\draftfalse

\journalname{Journal of Geophysical Research: Space Physics}

\begin{document}

%%%%%%%%%%%%%%%%%%%%%%%%%%%%%%%%%%%%%%%%%%%%%%%
%  TITLE
%
% (A title should be specific, informative, and brief. Use
% abbreviations only if they are defined in the abstract. Titles that
% start with general keywords then specific terms are optimized in
% searches)
%
%%%%%%%%%%%%%%%%%%%%%%%%%%%%%%%%%%%%%%%%%%%%%%%

% Example: \title{This is a test title}

\title{Capturing Secondary Kinetic Instabilities in Three-Dimensional Dayside Reconnection Using an Improved Gradient-Based Closure}

%%%%%%%%%%%%%%%%%%%%%%%%%%%%%%%%%%%%%%%%%%%%%%%
%
%  AUTHORS AND AFFILIATIONS
%
%%%%%%%%%%%%%%%%%%%%%%%%%%%%%%%%%%%%%%%%%%%%%%%

% Authors are individuals who have significantly contributed to the
% research and preparation of the article. Group authors are allowed, if
% each author in the group is separately identified in an appendix.)

% List authors by first name or initial followed by last name and
% separated by commas. Use \affil{} to number affiliations, and
% \thanks{} for author notes.
% Additional author notes should be indicated with \thanks{} (for
% example, for current addresses).

% Example: \authors{A. B. Author\affil{1}\thanks{Current address, Antartica}, B. C. Author\affil{2,3}, and D. E.
% Author\affil{3,4}\thanks{Also funded by Monsanto.}}

\authors{K. Bradshaw\affil{1}, A. H. Hakim\affil{2}, J. Juno\affil{2}, J. Pawlak\affil{1,2}, J. M. TenBarge\affil{1}, and A. Bhattacharjee\affil{1}}

% \affiliation{1}{First Affiliation}
% \affiliation{2}{Second Affiliation}
% \affiliation{3}{Third Affiliation}
% \affiliation{4}{Fourth Affiliation}

\affiliation{1}{Department of Astrophysical Sciences, Princeton University, Princeton, NJ, USA}
\affiliation{2}{Princeton Plasma Physics Laboratory, Princeton, NJ, USA}
%(repeat as many times as is necessary)

% Corresponding author mailing address and e-mail address:

% (include name and email addresses of the corresponding author.  More
% than one corresponding author is allowed in this LaTeX file and for
% publication; but only one corresponding author is allowed in our
% editorial system.)

% Example: \correspondingauthor{First and Last Name}{email@address.edu}

\correspondingauthor{K. Bradshaw}{kb8689@princeton.edu}

%%%%%%%%%%%%%%%%%%%%%%%%%%%%%%%%%%%%%%%%%%%%%%%
% KEY POINTS
%%%%%%%%%%%%%%%%%%%%%%%%%%%%%%%%%%%%%%%%%%%%%%%
%  List up to three key points (at least one is required)
%  Key Points summarize the main points and conclusions of the article
%  Each must be 140 characters or fewer with no special characters or punctuation and must be complete sentences

% Example:
% \begin{keypoints}
% \item	List up to three key points (at least one is required)
% \item	Key Points summarize the main points and conclusions of the article
% \item	Each must be 140 characters or fewer with no special characters or punctuation and must be complete sentences
% \end{keypoints}

\begin{keypoints}
\item An improved gradient-based heat flux closure is applied to multifluid simulations of the MMS Burch event.
\item Better agreement with kinetic results is achieved, including capture of current sheet instabilities and subsequent turbulence not replicated with prior closures.
\item The study demonstrates the ability of the ten-moment system to accurately capture kinetic instabilities driven by reconnection.
\end{keypoints}

%%%%%%%%%%%%%%%%%%%%%%%%%%%%%%%%%%%%%%%%%%%%%%%
%
%  ABSTRACT and PLAIN LANGUAGE SUMMARY
%
% A good Abstract will begin with a short description of the problem
% being addressed, briefly describe the new data or analyses, then
% briefly states the main conclusion(s) and how they are supported and
% uncertainties.

% The Plain Language Summary should be written for a broad audience,
% including journalists and the science-interested public, that will not have 
% a background in your field.
%
% A Plain Language Summary is required in GRL, JGR: Planets, JGR: Biogeosciences,
% JGR: Oceans, G-Cubed, Reviews of Geophysics, and JAMES.
% see http://sharingscience.agu.org/creating-plain-language-summary/)
%
%%%%%%%%%%%%%%%%%%%%%%%%%%%%%%%%%%%%%%%%%%%%%%%

%% \begin{abstract} starts the second page

\begin{abstract}
Magnetic reconnection is a highly dynamic process that excites a wide variety of kinetic waves and instabilities. Transverse current sheet instabilities such as the lower-hybrid drift and secondary drift-kink instabilities in particular have been shown by kinetic simulations to modify the reconnection and introduce significant turbulence and mixing to the reconnection layer. Past studies using the ten-moment fluid model to capture important kinetic physics such as the electron inertia and full representation of the pressure tensor proved advantageous to a two-fluid representation of reconnection, but the model struggled when using a local relaxation closure for the heat flux to replicate the current sheet instabilities and subsequent mixing seen in kinetic simulations. This work uses the \texttt{Gkeyll} software framework to perform simulations of asymmetric reconnection based on the 16 October 2015 MMS crossing of a diffusion region, the Burch event. An improved gradient-based heat flux closure is implemented, showing significant improvement in secondary kinetic instabilities that grow in the current sheet. These instabilities generate turbulence which leads to growth of secondary magnetic islands and flux ropes.
\end{abstract}

%%%%%%%%%%%%%%%%%%%%%%%%%%%%%%%%%%%%%%%%%%%%%%%
%
%  BODY TEXT
%
%%%%%%%%%%%%%%%%%%%%%%%%%%%%%%%%%%%%%%%%%%%%%%%

%%% Suggested section heads:
% \section{Introduction}
%
% The main text should start with an introduction. Except for short
% manuscripts (such as comments and replies), the text should be divided
% into sections, each with its own heading.

% Headings should be sentence fragments and do not begin with a
% lowercase letter or number. Examples of good headings are:

% \section{Materials and Methods}
% Here is text on Materials and Methods.
%
% \subsection{A descriptive heading about methods}
% More about Methods.
%
% \section{Data} (Or section title might be a descriptive heading about data)
%
% \section{Results} (Or section title might be a descriptive heading about the
% results)
%
% \section{Conclusions}

\section{Introduction}

The planetary magnetosphere plays a key role in the shielding of the Earth from the energetic solar wind. Improved modeling of the coupling between the Earth's magnetosphere and the solar wind is of great importance, particularly in understanding and predicting the space weather events which can impact technological infrastructure \cite{baker2004}. One frequent phenomenon that occurs within the magnetosphere is magnetic reconnection. The primary locations at which reconnection occur are dayside near the heliopause, and nightside in the magnetosphere tail. In dayside reconnection regions where the solar wind magnetic fields stream in the opposite direction of the dipole field, reconnection occurs frequently and results in energy transfer from the solar wind into the magnetosphere \cite{paschmann1979plasma}. The Magnetospheric Multiscale (MMS) mission has provided access to high-resolution in situ measurements of the of the dissipation region where this energy transfer occurs. The first such set of measurements was the 16 October 2015 crossing \cite{burch2016electron}, referred to hereafter as the Burch event. This data is an invaluable resource to simulation frameworks as it provides an opportunity to test the fidelity of the physics being captured by the underlying models.\par
Traditional modeling of the magnetosphere has relied predominantly on a confluence of magnetohydrodynamics (MHD), hybrid kinetic, and particle-in-cell methods, each with strengths and limitations. MHD, while reproducing global dynamics surprisingly well, fails near the dissipation region since magnetospheric plasmas are typically collisionless \cite{hesse2011diffusion}. Resistive Hall MHD does better in the reconnection boundary-layers \cite{birn2001geospace,liu2025ohm}, but the assumption of scalar pressure is flawed since the pressure tensor and electron inertia have been found to be important, theoretically as well as observationally \cite{liu2025ohm}. Hybrid kinetic models typically evolve the Vlasov equation for ions but use fluid electrons due to how restrictive electron kinetic scales are to resolve. This approach has similar issues to expanded MHD methods, failing to capture important electron dynamics contained within the full pressure tensor. More sophisticated electron models have been utilized on top of existing hybrid kinetic frameworks \cite{omelchenko2012,omelchenko2021,omelchenko2021b}, assuming the magnetic field evolution comes from the hybrid kinetic simulation \cite{battarbee2021,ganse2023}.\par
MHD and more recently with improved algorithms and computational power, hybrid kinetic, have been used extensively for global modeling of planetary magnetospheres \cite{lavraud2008altered,pokhotelov2013,palmroth2018vlasov,burkholder2024global}. An alternative approach to global magnetosphere modeling which seeks to add the missing electron physics is multi-moment multifluid models. This approach has been applied to a variety of systems, including Ganymede \cite{wang2018electron}, Mercury \cite{dong2019global}, and Earth \cite{wang2020exact}. The addition of the full pressure tensor evolution allows for improved physics fidelity of reconnection simulations, comparing favorably to kinetic simulations for key metrics such as the Ohm's Law \cite{wang2015comparison} and scaling of the reconnection rate with island size \cite{ng2015island}. Higher moment multifluid models allow for great improvements in capturing kinetic physics over MHD models, especially in regions where electron inertia and pressure agyrotropy drive electron demagnetization. The focus of this work is the ten-moment model introduced by \citeA{wang2015comparison}, which retains finite mass and includes the evolution of the full pressure tensor.\par
Simulations of magnetic reconnection which retain finite electron inertia have revealed a number of instabilities which grow in the reconnection region and influence the reconnection properties \cite{graham2025role}. Fluid and hybrid simulations often struggle to replicate all of these phenomena due to lacking the aforementioned electron dynamics. Past multifluid simulations performed by \citeA{tenbarge2019extended} using the \texttt{Gkeyll} software framework were capable of replicating many of the important features of the MMS data and kinetic simulations, but there were several prominent features which were not captured by the model. The most notable of these is the growth of the lower-hybrid drift instability (LHDI) in the transverse direction along the current sheet \cite{yoon2002generalized,roytershteyn2012influence}. LHDI is predicted by kinetic simulations to grow rapidly during reconnection, and to lead to increased turbulence and mixing of electrons between the layers \cite{le2017enhanced}. The limiting factor to the effectiveness of this previous work was the use of an overly simplistic heat flux closure for reconnection. The approximation, based on the work of \citeA{hammett1990fluid}, took the form of a local relaxation of the pressure tensor towards isotropy. There has been significant effort to improve closures for the full pressure tensor in the intervening years, with both nonlocal and gradient-based generalizations of this relaxation closure \cite{allmannrahn2018,ng2020improved,allmannrahn2021,kuldinow2024}. In particular, the gradient-based approaches that retain the form of a relaxation operator, but now acting on all $k$'s instead of just a single $k_0$, have shown promise for capturing the very instabilities which the local closure struggles to model in these parameter regimes.\par
This paper shows that limitations with the previous work can be addressed by utilizing a gradient-based closure previously developed by \citeA{ng2020improved} as a generalization of a diffusive heat-flux closure to the ten-moment system. This approach can be contrasted with these relaxation operators in that it seeks to capture the actual thermal transport of the plasma, with the potential for the transport to be modified in each direction independently in this tensorial generalization of Fick’s Law, exactly as we expect in a plasma system in which the flow of heat will vary due to, e.g., the local magnetic field direction. The work presented here marks a notable improvement on past multifluid simulations of magnetic reconnection in the magnetosphere. It demonstrates the ability of the ten-moment model to capture kinetic physics in regimes previously inaccessible to fluid solvers. Section~\ref{sec:model} gives an overview of the \texttt{Gkeyll} software framework, the ten-moment model and closure, and numerical method including heat flux limiters. Section~\ref{sec:simulation} provides the simulation initial conditions and results. Section~\ref{sec:discussion} gives discussion of these results with an eye to physics fidelity. Finally, a summary of the paper and grounds for future work are provided in Section~\ref{sec:conclusion}.\par

\section{The Ten-Moment Two-Fluid Model}\label{sec:model}

\subsection{System of Equations}

The ten-moment model \cite{wang2015comparison} may be derived by taking moments up to second order of the Vlasov equation,
\begin{equation}
    \frac{\partial n}{\partial t} + \frac{\partial nu_i}{\partial x_i} = 0,
\end{equation}
\begin{equation}
    m\frac{\partial nu_i}{\partial t} + \frac{\partial\mathcal{P}_{ij}}{\partial x_j} = nq(E_i + \epsilon_{ijk}u_jB_k),
\end{equation}
\begin{equation}
    \frac{\partial\mathcal{P}_{ij}}{\partial t} + \frac{\partial\mathcal{Q}_{ijk}}{\partial x_k} = nqu_{[i}E_{j]} + \frac{q}{m}\epsilon_{[ikl}\mathcal{P}_{kj]}B_l.
\end{equation}
Square brackets indicate the sum over symmetric index permutations, i.e. $u_{[i}E_{j]} = u_iE_j + u_jE_i$. The electromagnetic fields are coupled to the system through Maxwell's equations,
\begin{equation}
	\frac{\partial B_i}{\partial t} = -\epsilon_{ijk}\frac{\partial E_k}{\partial x_j},
\end{equation}
\begin{equation}
	\frac{1}{\mu_0\varepsilon_0}\frac{\partial E_i}{\partial t} + \mu_0 J_i = \epsilon_{ijk}\frac{\partial B_k}{\partial x_j}.
\end{equation}
This system of equation makes use of the following high-order moments of the distribution function,
\begin{equation}
    \mathcal{P}_{ij} = m\int v_iv_jfd\mathbf{v},
\end{equation}
\begin{equation}
    \mathcal{Q}_{ijk} = m\int v_iv_jv_kfd\mathbf{v},
\end{equation}
where $d\mathbf{v}$ denotes an integration over all three velocity dimensions. The second moment may be rewritten in terms of the pressure tensor by the relation 
\begin{equation}
    \mathcal{P}_{ij} = P_{ij} + nmu_iu_j,
\end{equation}
while the relationship between the third moment and the heat flux tensor may be written as
\begin{equation}
    \mathcal{Q}_{ijk} = Q_{ijk} + u_{[i}\mathcal{P}_{jk]} - 2nmu_iu_ju_k.
\end{equation}
The total ten-moment system is ten equations (1 continuity + 3 momentum + 6 unique components of the pressure tensor), with ten additional unknowns from the unique components of the heat flux tensor. A closure is thus required for the heat flux tensor to complete the system. Previous publications on magnetic reconnection with the \texttt{Gkeyll} multifluid app have primarily used the \textit{local closure} modified from \citeA{hammett1990fluid}, which is an approximation of the plasma response function based on a three-pole Pad\'{e} series.
\begin{equation}
    \frac{\partial Q_{ijk}}{\partial x_k} = -v_t|k_0|(P_{ij} - p\delta_{ij}),
\end{equation}
where $k_0$ is a constant diffusion parameter, and $v_t = \sqrt{T/m}$ is the thermal speed. This closure acts as a collision-like operator, relaxing the pressure temperature everywhere towards isotropy. This approach has the effect of damping out any developing anisotropic and agyrotropic features in the pressure tensor, which are expected in the dissipation region.\par
To better capture anisotropy and agyrotropy in the pressure tensor, this work instead favors the \textit{gradient closure} introduced by \citeA{ng2020improved}, based on \citeA{sharma2007preserving},
\begin{equation}
    Q_{ijk} = -\frac{v_t}{|k_0|}\partial_{[k}T_{ij]}.\label{eq:grad_closure}
\end{equation}
This closure ensures that heat flow is greatest in directions where there is a large temperature gradient, without isotropizing the pressure tensor. This approach differs from the gradient closure used by \citeA{allmannrahn2018} and \citeA{kuldinow2024} in that it acts on the temperature instead of the pressure, preserving a form similar to the standard Fick's law.\par

\subsection{Numerical Method}

\texttt{Gkeyll} is a plasma physics software framework including continuum kinetic Vlasov and gyrokinetic solvers, as well as multifluid five and ten-moment finite volume solvers. For the fluid solvers, the code employs a finite volume high-resolution wave propagation method \cite{leveque2002finite}, following the algorithm detailed by
\citeA{hakim2006high}, \citeA{hakim2008extended}, and \citeA{loverich2013nautilus}. The update is split between the homogenous part of the equation and the source terms which couple the fluid and field equations, with a second-order Strang-splitting algorithm being used for the time integration \cite{strang1968construction}. In the homogeneous update, Roe averaging \cite{roe1981approximate} is used at the cell interfaces to solve the Riemann problem, unless positivity of the density or pressure is violated, which prompts a recomputation of the update using Lax fluxes. The source terms are updated using a locally implicit algorithm \cite{wang2020exact}, which eliminates restrictions due to kinetic scales and allows for a less constrained time step.\par
The original implementation of the gradient closure by \citeA{ng2020improved} has been modified algorithmically to follow a symmetric heat diffusion scheme \cite{gunter2005modelling,gunter2007finite} which evaluates temperature and heat fluxes at the vertices (see \ref{app:closure}). The closure being evaluated at the vertices can lead to unphysical directions of heat flow when adjacent cells have flow in different directions. To prevent flow from cold regions to hot regions (which can cause positivity issues), the symmetric limiter proposed by \citeA{sharma2007preserving} is implemented.\par

\section{MMS Reconnection Simulations}\label{sec:simulation}

\subsection{Initial Conditions}

The initial conditions are based on the conditions present during the 2016 Burch event \cite{burch2016electron} crossing of a reconnection dissipation region. To satisfy periodic boundary conditions, a symmetric dual current sheet is initialized. Profiles for the initial temperature and magnetic field of the sheet are

\begin{equation}
    f(Q_s,Q_m) = 0.5[Q_s + Q_m + (Q_m - Q_s)\tanh{(y/w_0)}],
\end{equation}
where $w_0 = d_{is}$ is the current sheet width, and $Q_s$ and $Q_m$ are the values of the corresponding quantities in their respective regions. $d_{is}$ is the asymptotic ion inertial length, defined in terms of the ion plasma frequency $\omega_{pis}$ as $d_{is} = c/\omega_{pis}$. The subscript $s$ is used to denote quantities in the magnetosheath, and $m$ to denote quantities in the magnetosphere. Temperatures used are  $T_{em}/T_{es} = 12.3$ and $T_{im}/T_{is} = 5.63$, while the density ratio is $n_m/n_s = 0.062$. The initial magnetic field is $B_{xm}/B_{xs} = -1.696$. For ease of notation in the rest of this work, we will take $d_i$ to be $d_{is}$ and the ion cyclotron frequency $\Omega_{ci}$ to be $\Omega_{cis}$\par
The simulation resolution used is $[N_x, N_y, N_z] = [1152, 576, 288]$, covering a domain length of $[L_x, L_y, L_z] = [40.96d_i, 20.48d_i, 10.24d_i]$ with fully periodic boundaries in each direction. The mass fraction is set to a reduced value of $m_i/m_e = 100$, while the asymptotic Alfv\'{e}n speed is reduced to $c/v_{A\perp s} = 200$, where $v_{A\perp s} = B_{xs}/\sqrt{\mu_0m_in_s}$. Previous studies of magnetic reconnection with \texttt{Gkeyll} performed by \citeA{ng2015island} and \citeA{wang2015comparison} demonstrated that a closure parameter $k_{0,\alpha}d_{\alpha}\sim1$, where $\alpha$ indicates species, gave the best results in comparison to PIC simulations, so that is the choice used here for both the local and gradient closure. A uniform guide field $B_z/B_{sx} = 0.099$ is used. The first twenty Fourier modes of $B_x$ and $B_y$ are initialized with noise at a root-mean-square amplitude $B_{noise} = 0.0005B_{sx}$. A perturbation is added to the vector potential
\begin{equation}
    A_z = \psi_0\cos{\bigg(\frac{2\pi x}{L_x}\bigg)}\bigg[1 - \cos{\bigg(\frac{4\pi y}{L_y}\bigg)}\bigg],
\end{equation}
with $\psi_0 = 0.1$. These modes are introduced to break the initial symmetry of the simulation \cite{tenbarge2014collisionless}.\par
The primary simulation of interest for this work uses the gradient closure. However, a local closure simulation with equivalent initial conditions was also run and is used for comparison for several of the key results in the next section.\par

\subsection{Results}

\begin{figure}
    \includegraphics[width=1.0\linewidth]{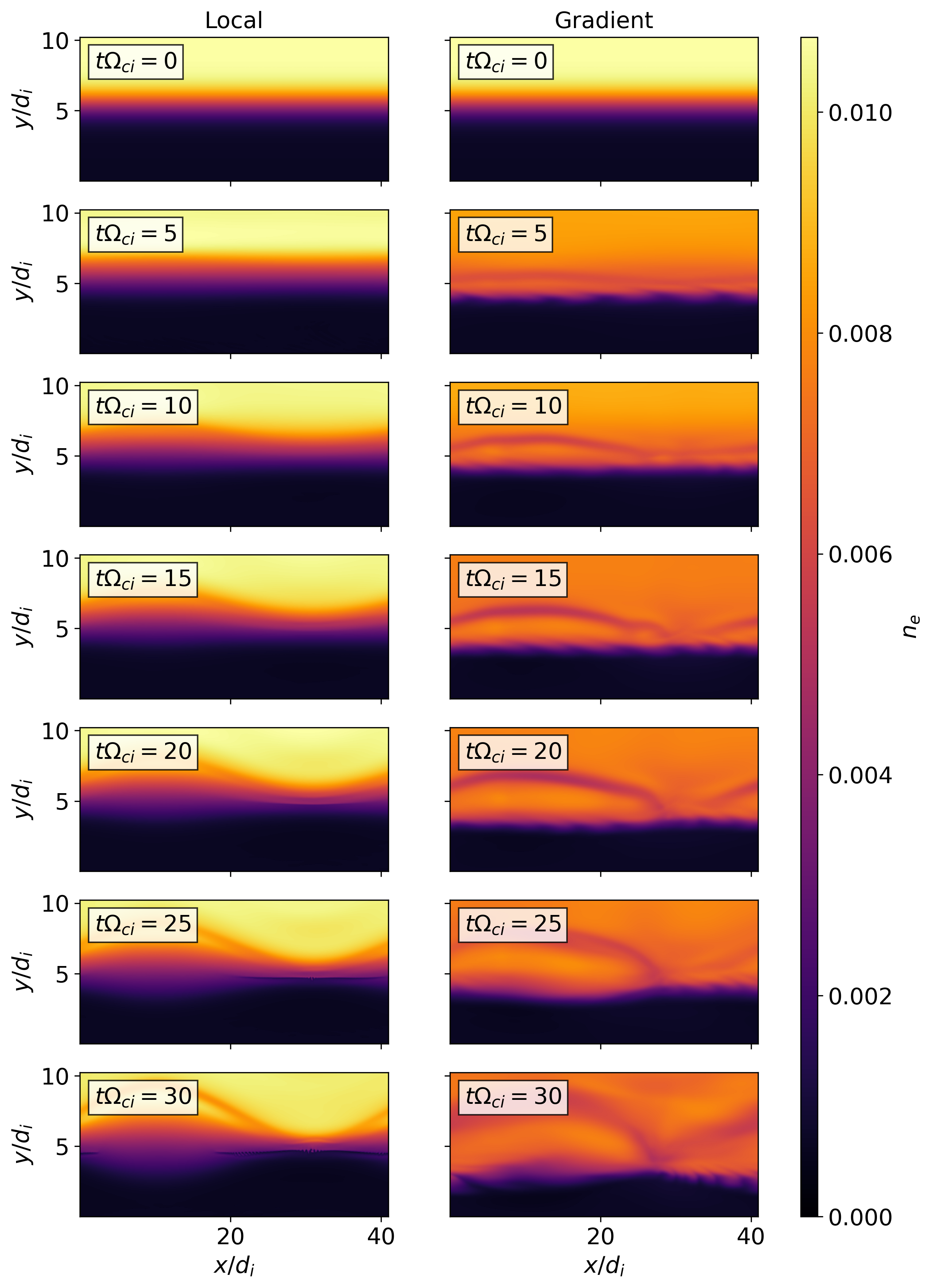}
    \caption{\label{fig:burch_elc} Electron density snapshots at $z=L_z/2$. The electrons are significantly more diffuse in the gradient closure case (right), and the reconnection layer itself has an island with a density plateau, while in the local closure simulation (left) the layer has a smooth density gradient across the island.}
\end{figure}

\begin{figure}
    \includegraphics[width=1.0\linewidth]{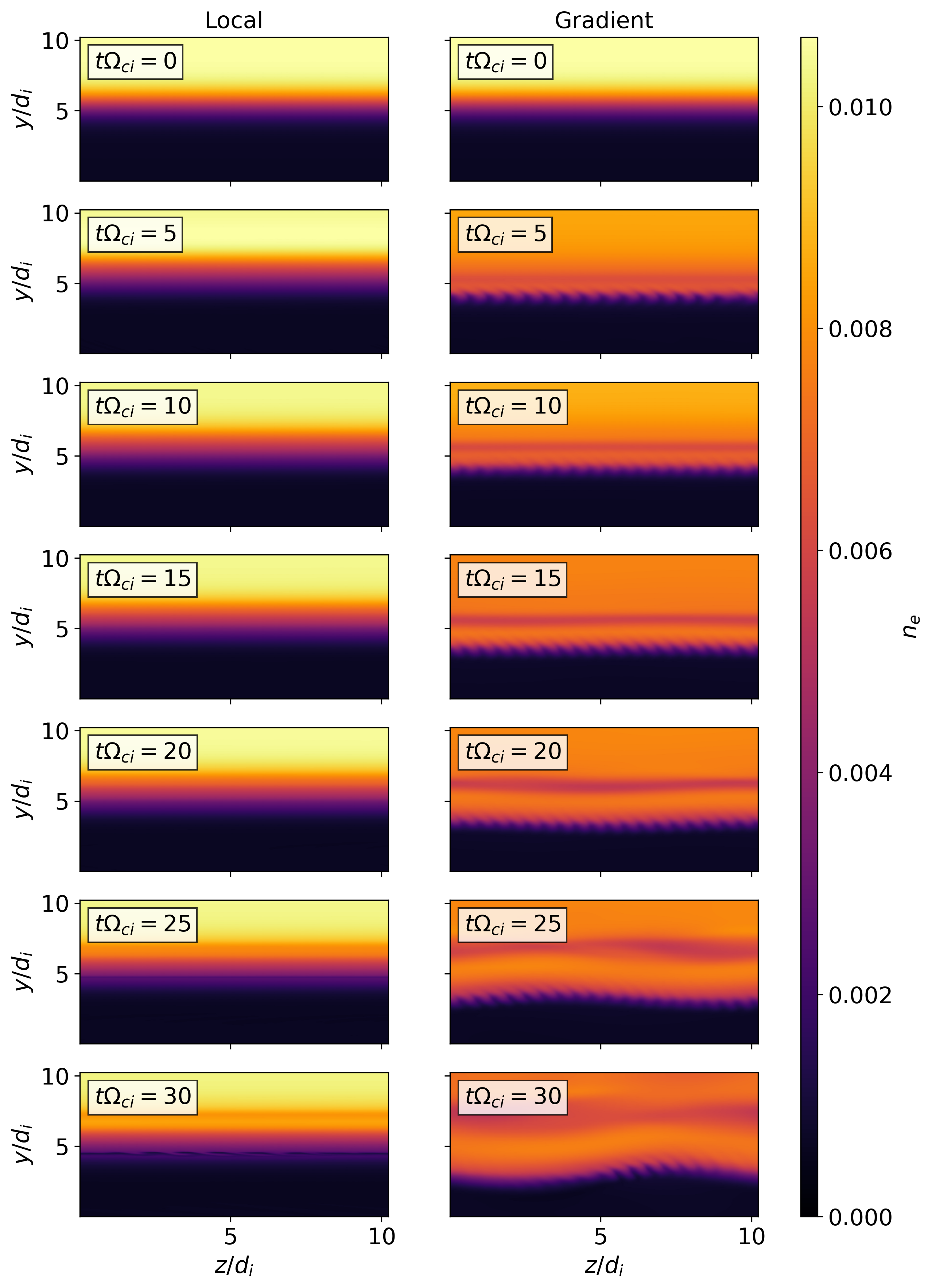}
    \caption{\label{fig:burch_lhdi} Slices of the electron density taken at the X-line across various times. No significant mode growth occurs for the simulation run using the local closure (left), while the gradient closure (right) allows LHDI growth at the edge of the current sheet. During nonlinear saturation, the LHDI triggers a shearing instability.}
\end{figure}

\begin{figure}
    \centering
    \includegraphics[width=0.49\linewidth]{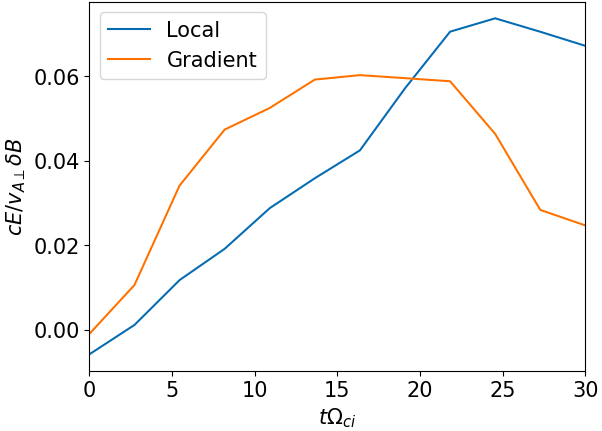}
    \includegraphics[width=0.49\linewidth]{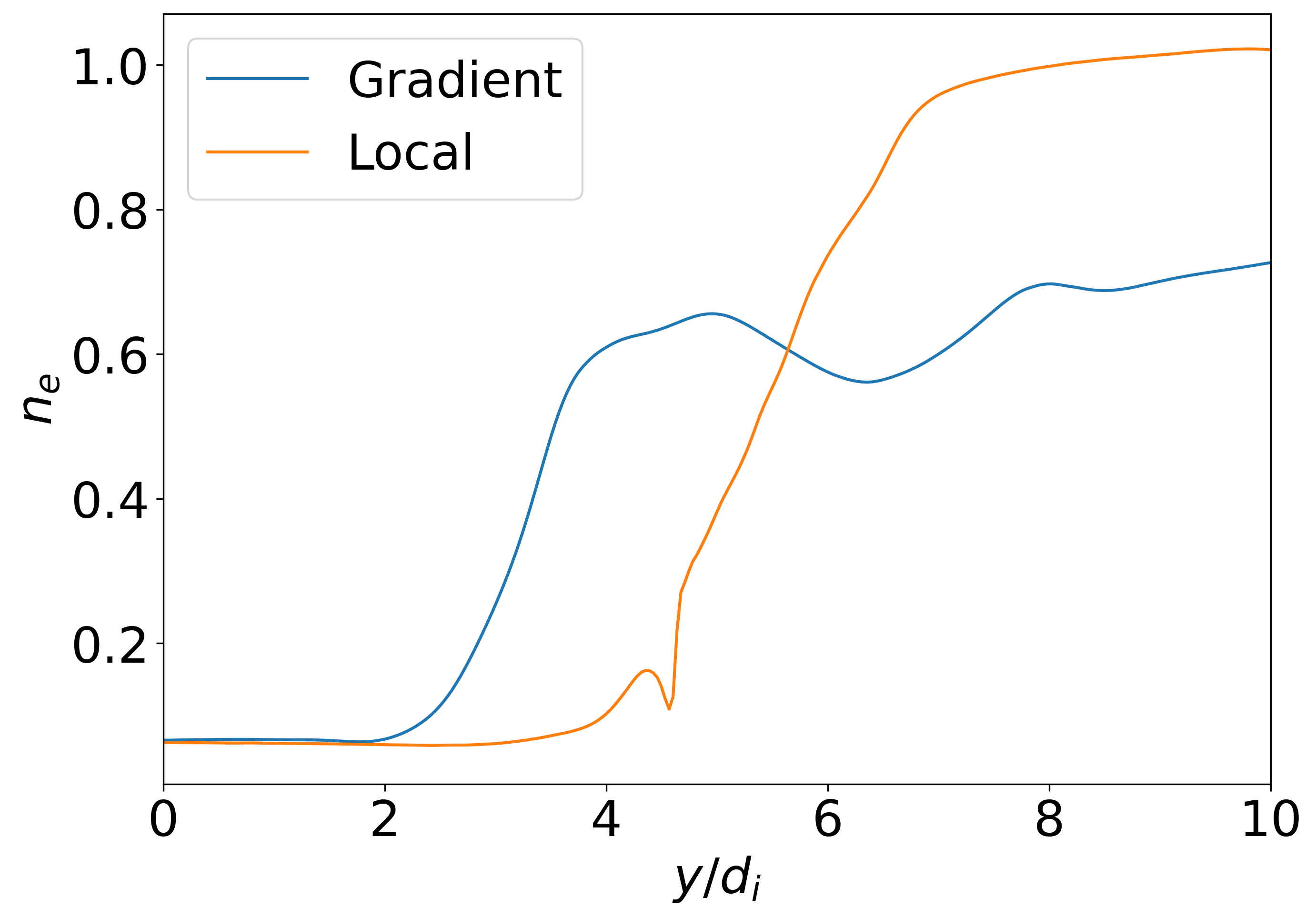}
    \caption{\label{fig:recon_rate} Magnetic reconnection rates (left), and density gradients (right) across the reconnection layer and separatrices at $x=20d_i$ and averaged over the $z$-direction. The gradient closure experiences a steady decline in the reconnection rate around the saturation time of the LHDI. The decreasing density profile across the separatrices in the local closure results demonstrates significant diffusion perpendicular to the field lines, while the plateaus and sharper drops in the gradient closure suggest not much diffusion is occurring cross-field.}
\end{figure}

The evolution of the electron density for the two different closures is plotted in the reconnection plane in Fig.~\ref{fig:burch_elc} and the transverse plane in Fig.~\ref{fig:burch_lhdi}. The gradient closure gives densities which are more diffuse than the local closure, with a wider and more uniform layer. For the local closure, there is a steady density gradient across the layer declining from the magnetosheath side to the magnetosphere side (see the second plot in Fig.~\ref{fig:recon_rate}), while the gradient closure has a nearly flat profile across the layer, with a sharp transition to low density at the magnetosphere separatrix. The gradient closure is a significantly better match to PIC results for asymmetric reconnection \cite{yamada2018two}, which predict low rates of diffusion across field lines. Growth of some unstable modes is also visible along the low-density separatrix, likely driven by shear flow between the layers. In the transverse direction, the difference in physics captured by the two closures is even more pronounced. No instabilities grow for the local closure, but in the gradient closure case, LHDI begins growing within five ion cyclotron periods, and the mode ultimately saturates, causing a kink instability to begin forming at around $t\Omega_{ci} = 20$, a phenomenon observed in many previous kinetic simulations of the LHDI \cite{yoon2002generalized}.\par

Reconnection rates are plotted in Fig.~\ref{fig:recon_rate}. The rates are calculated by constructing a vector potential $\langle A_z\rangle$ at the X-point by averaging the in-plane magnetic fields over the $z$-direction, and using this average to construct the reconnection field by $\frac{\partial\langle A_z\rangle}{\partial t}$. Compared to the local closure, the gradient closure peaks earlier, and instead of plateauing begins to decrease steadily. This behavior is similar to that observed in the kinetic simulations performed by \citeA{le2017enhanced}, and the onset of the decline coincides approximately with the nonlinear saturation of the LHDI and onset of the kink instability. The reconnection rate of the gradient closure simulation also peaks somewhat lower than the canonically observed value of $cE/v_{A\perp s}B_{xs}\sim 0.1$.\par
\begin{figure}
    \includegraphics[width=1.0\linewidth]{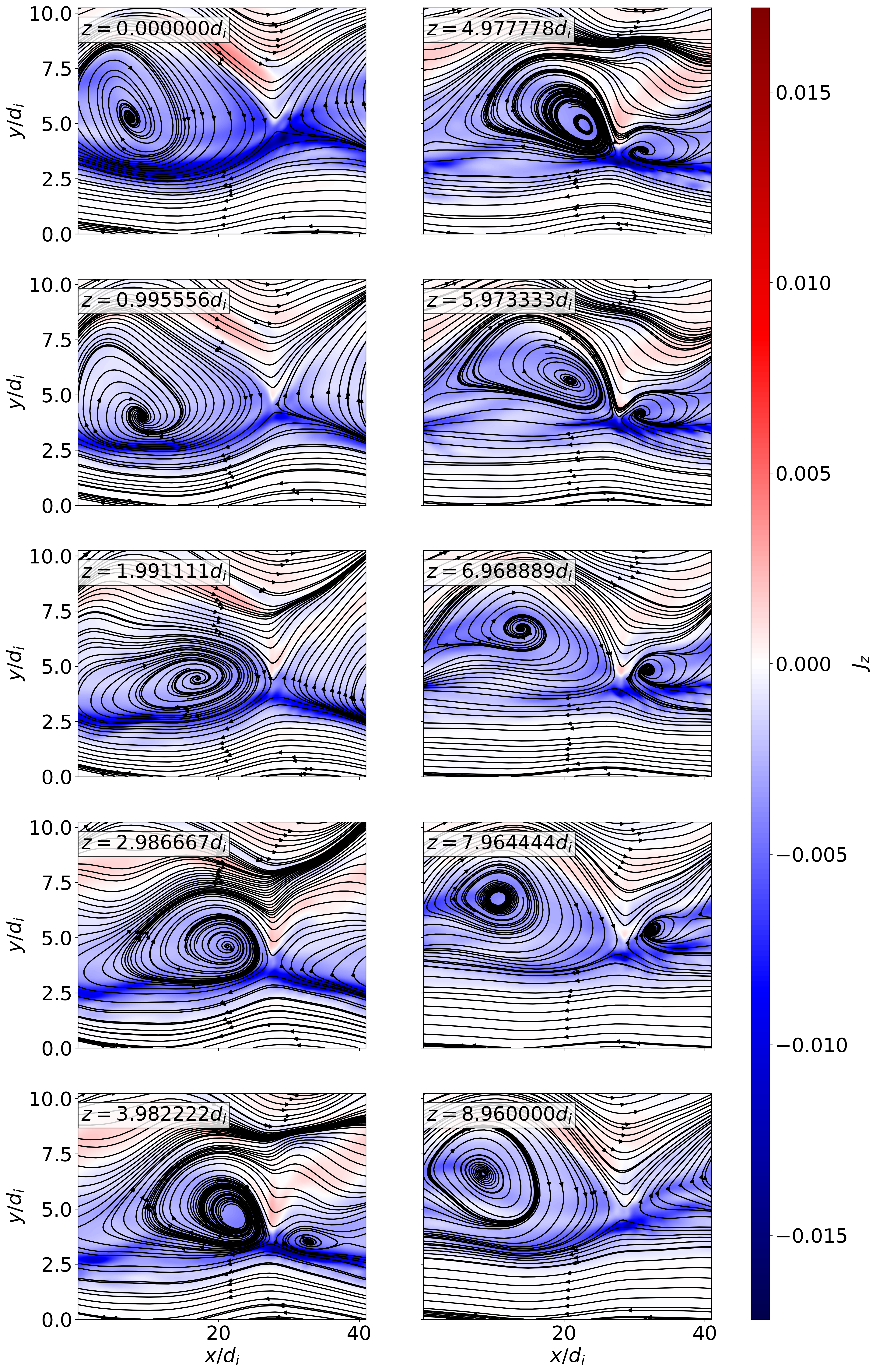}
    \caption{\label{fig:burch_layer} Transverse current $J_z$ (color), and in-plane magnetic field lines (black lines) in the reconnection plane of the gradient closure simulation. Significant distortion of the flux tube in the transverse direction occurs as the kink instability develops.}
\end{figure}
\begin{figure}
    \includegraphics[width=1.0\linewidth]{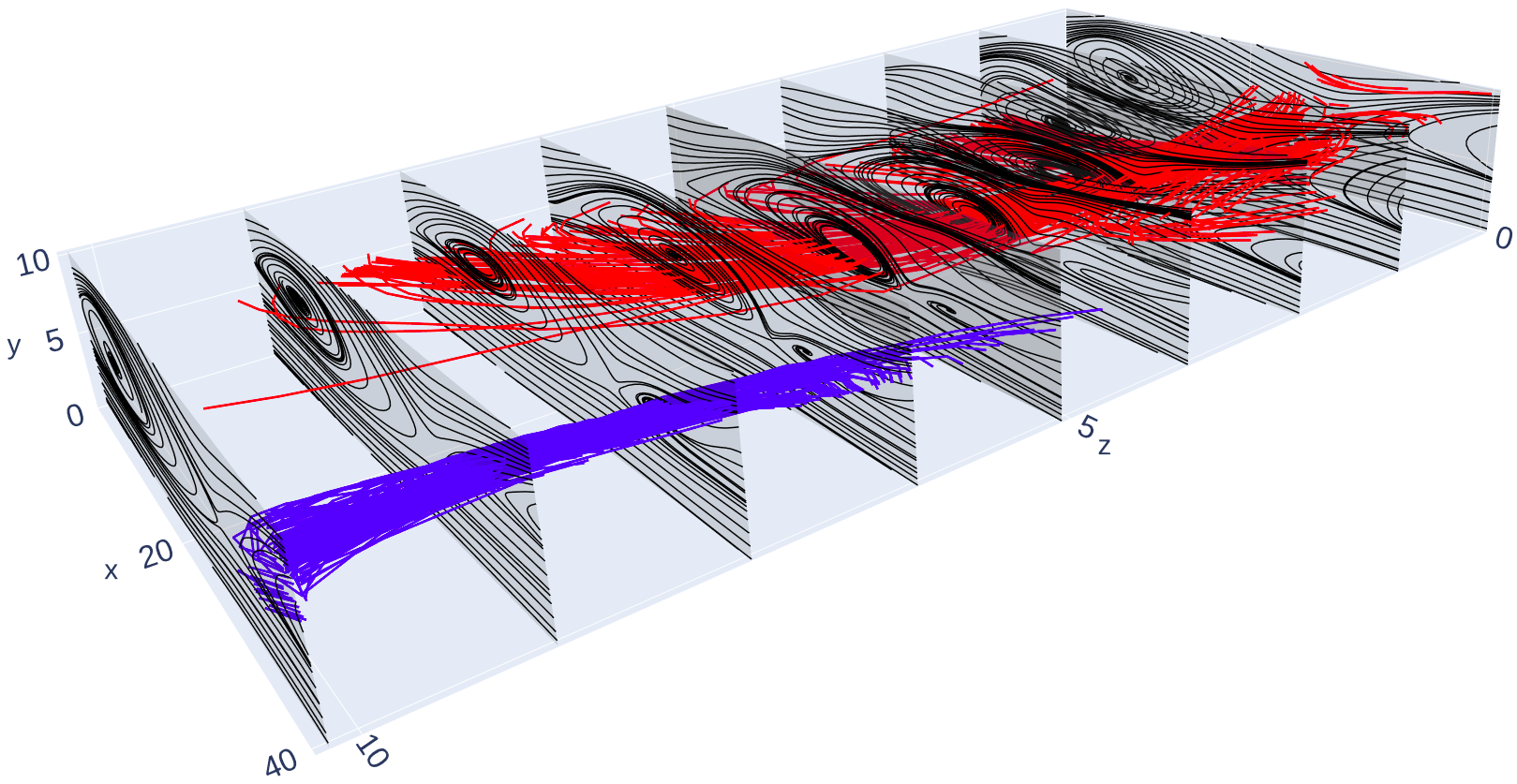}
    \caption{\label{fig:flux_ropes} 3D visualization of the magnetic field topology with field lines traced through regions of high magnetic flux for the gradient closure. The primary bundle of field lines (red, $+B_z$) passes directly through the layer, veering to the side and ultimately reversing direction near the edge of the domain. A secondary tube of field lines (blue, $-B_z$) extends the opposite direction through the island that forms.}
\end{figure}

Slices of the current density along the $z$-axis are plotted in Fig.~\ref{fig:burch_layer}, along with streamlines of the in-plane magnetic field. The layer is very nonuniform in $z$, with an island forming in one region. 3D visualization of the field line topology in Fig.~\ref{fig:flux_ropes} shows formation of flux ropes similar to behavior observed by \citeA{daughton2011role}. In particular, there is one major bundle of $+B_z$ field lines passing through the main layer, and a smaller bundle of $-B_z$ field lines passing through the island.\par

\begin{figure}
    \includegraphics[width=0.49\linewidth]{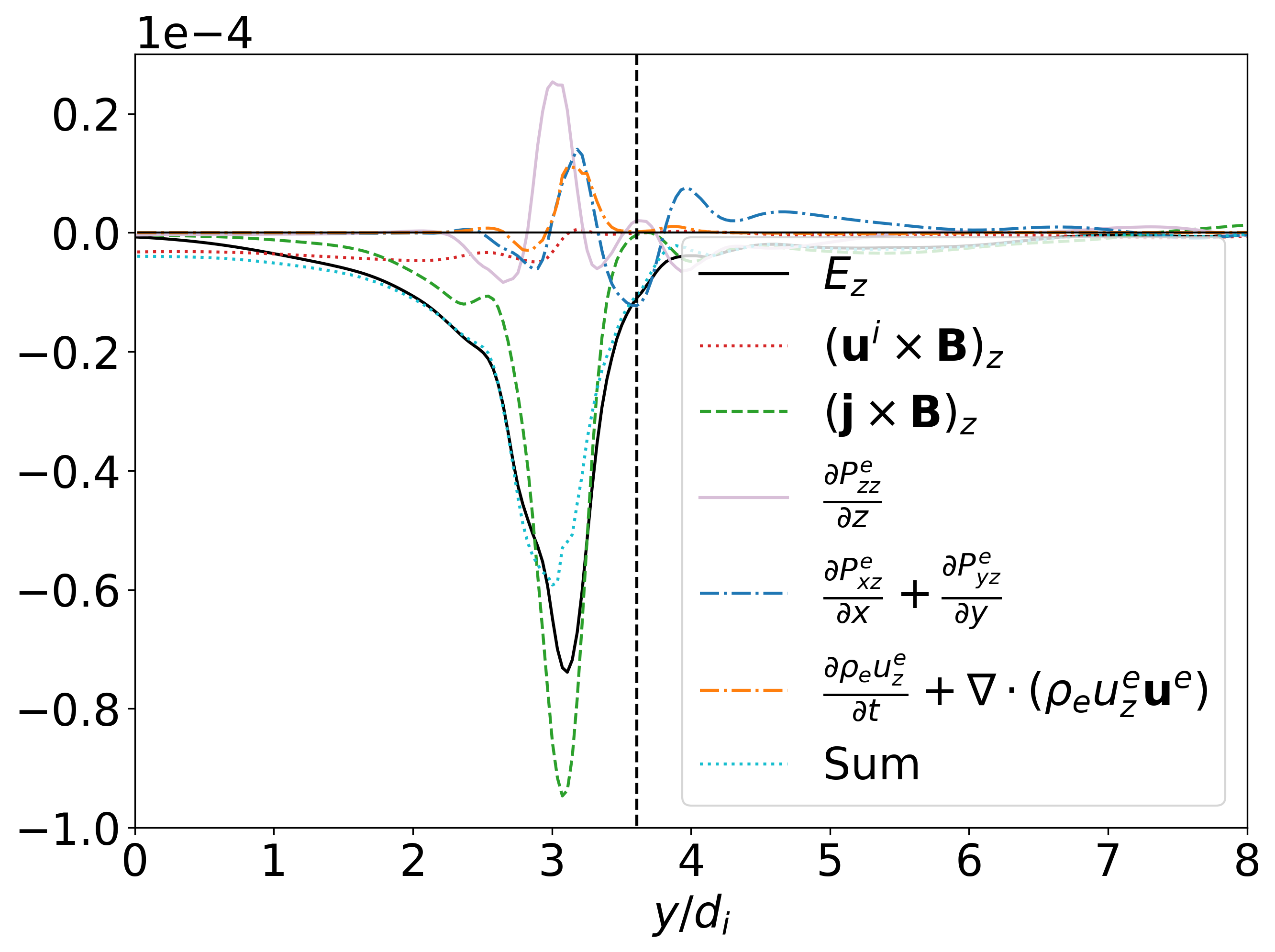}
    \includegraphics[width=0.49\linewidth]{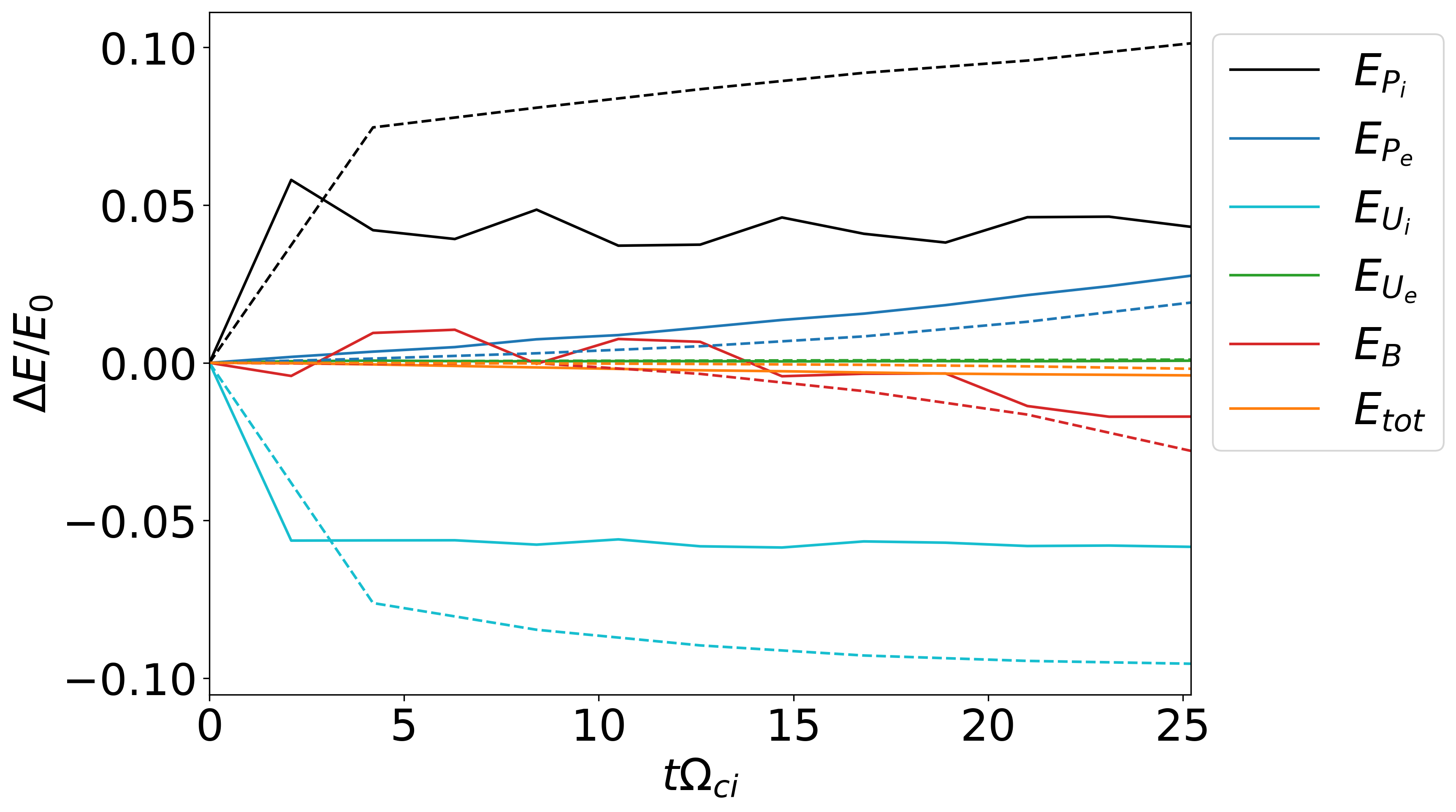}
    \caption{\label{fig:burch_energy} Components of the generalized Ohm's law (left) across the current layer at $z = L_z/2$ for the gradient closure. The vertical dashed line marks the location of the X-point. Near the X-point, only the pressure tensor terms are significant. Contributions to total energy (right) by different terms for the gradient (solid) and local (dashed) closure results. The gradient closure experiences less significant of a relative change in energy overall, with the ion pressure and flow energy exchange in particular being significantly smaller. The sole quantity which increases at a greater rate than with the local closure is the electron pressure.}
\end{figure}

Fig.~\ref{fig:burch_energy} gives the breakdown of the components of the generalized Ohm's law across the X-line, where only the component aligned with the guide field contributes,

\begin{equation}\label{eq:ani_closure}
    \begin{split}
        E_z = &-(\mathbf{u}^i\times\mathbf{B})_z + \frac{1}{n_e|e|}(\mathbf{j}\times\mathbf{B})_z - \frac{1}{n_e|e|}\bigg(\frac{\partial P_{xz}^e}{\partial x} + \frac{\partial P^e_{yz}}{\partial y} + \frac{\partial P^e_{zz}}{\partial z}\bigg)\\
        &- \frac{1}{n_e|e|}\bigg[\frac{\partial\rho_eu_z^e}{\partial t} + \nabla\cdot(\rho_eu_z^e\mathbf{u}^e)\bigg].
    \end{split}
\end{equation}

This equation, while not directly represented in the underlying equation system, but inferred from it, gives the representation of the breaking of the frozen-in condition in the dissipation region. Directly at the X-point, only the pressure tensor terms are significant, with an additional spike of the Hall term across the reconnection layer. These features match well to kinetic results \cite{liu2014dispersive,le2017enhanced}. Fig.~\ref{fig:burch_energy} also gives a breakdown of the contributions to the total energy over the volume $\mathcal{V}$ from the magnetic field ($E_B = \frac{1}{2\mu_0}\int_{\mathcal{V}} \mathbf{B}\cdot\mathbf{B}d\mathcal{V}$), drift ($E_U = \frac{1}{2}\int_{\mathcal{V}}\rho\mathbf{u}\cdot\mathbf{u}d\mathcal{V}$), and pressure ($U_P = \frac{1}{2}\int_{\mathcal{V}}\mathcal{P}_{ii}d\mathcal{V} - E_U$) terms, with comparison to the local closure. A relatively high amount of energy is contained within the electron pressure tensor in the gradient closure case.\par

\begin{figure}
    \includegraphics[width=1.0\linewidth]{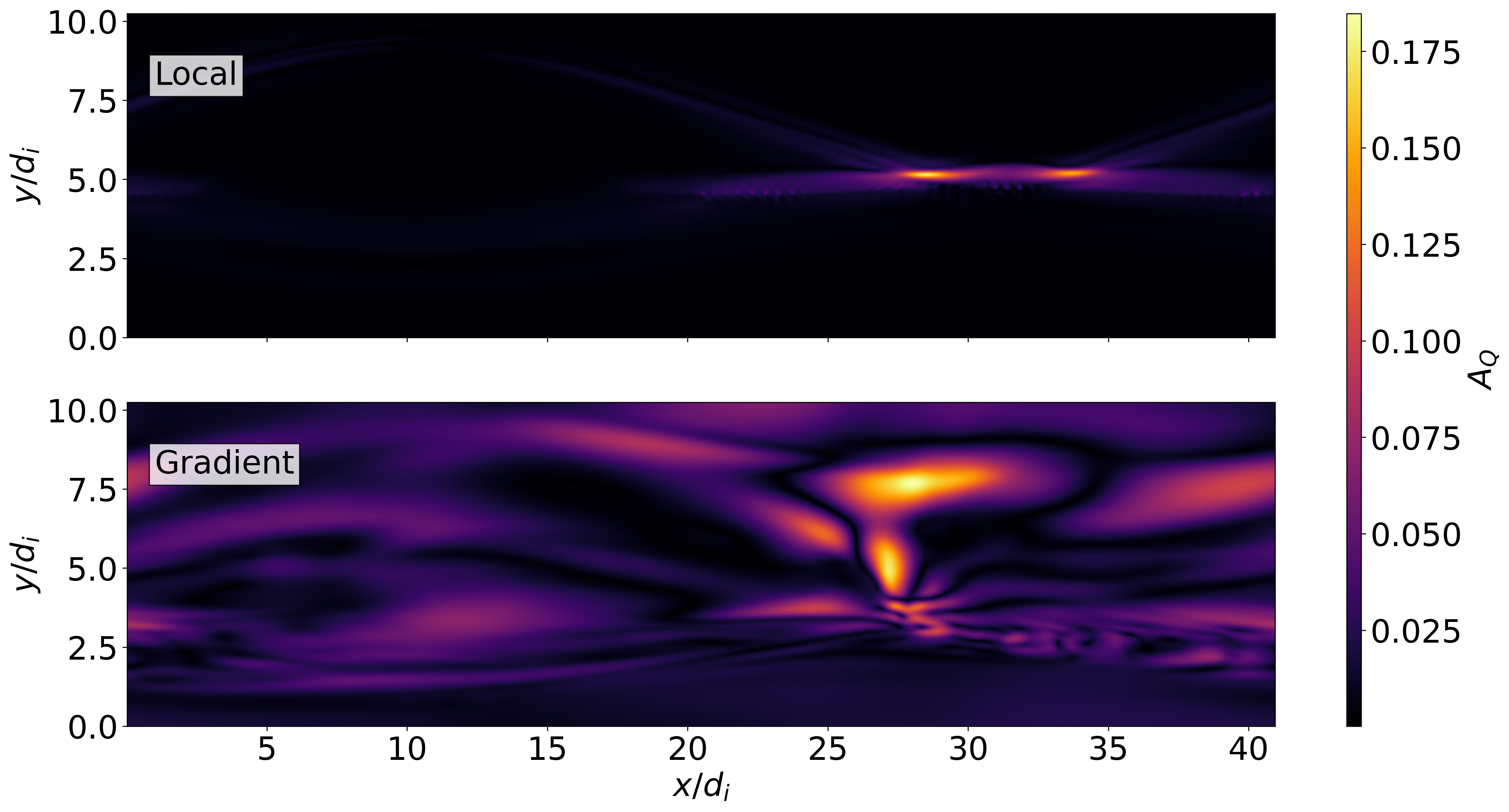}
    \caption{\label{fig:burch_agyro} Agyrotropy in the local (top) and gradient (bottom) closure simulations at $t\Omega_{ci} = 30$ and $z = L_z/2$. With the gradient closure, agyrotropy is not confined to the X-point and separatrices and extends to a greater extent into the layer and magnetosheath.}
\end{figure}

Continuing the emphasis on the pressure tensor, comparisons of the electron agyrotropy between the local and gradient closure are shown in Fig.~\ref{fig:burch_agyro}, where the \citeA{swisdak2016quantifying} measure of agyrotropy (here denoted $A_Q$ instead of $\sqrt{Q}$ to distinguish it from heat flux) is used:
\begin{equation}
    A_{Q} = \sqrt{\frac{P_{12}^2 + P_{13}^2 + P_{23}^2}{P_{\perp}^2 + 2P_{\perp}P_{\parallel}}}.
\end{equation}
Both closures produce strong agyrotropy near the X-point, as expected. With the gradient closure, this agyrotropy extends into the layer and even into the magnetosheath. Strong agyrotropy along the separatrices with the gradient closure matches well to kinetic results and is an improvement on what is seen with the local closure, but the overall agyrotropy region is significantly more expansive than is typically observed in reconnection \cite{swisdak2016quantifying,che2018quantifying} and likely occurs due to the lack of any closure-related isotropizing mechanism acting on the pressure tensor combined with the tendency of the gradient closure to exacerbate any existing anisotropies.

\begin{figure}
    \includegraphics[width=1.0\linewidth]{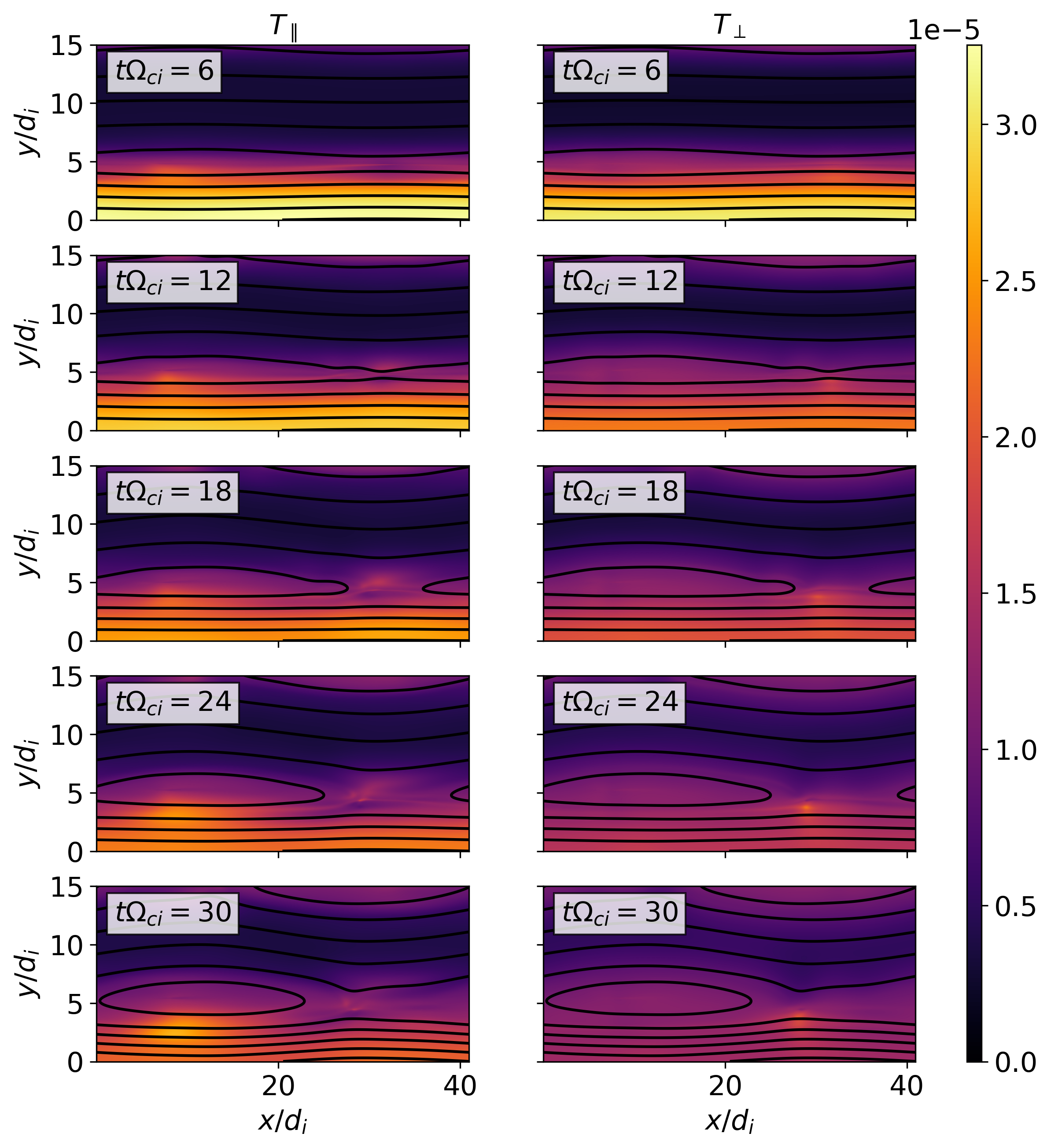}
    \caption{\label{fig:burch_temp} Parallel (left) and perpendicular (right) temperature in the reconnection layer for the gradient closure.}
\end{figure}

\begin{figure}
    \includegraphics[width=1.0\linewidth]{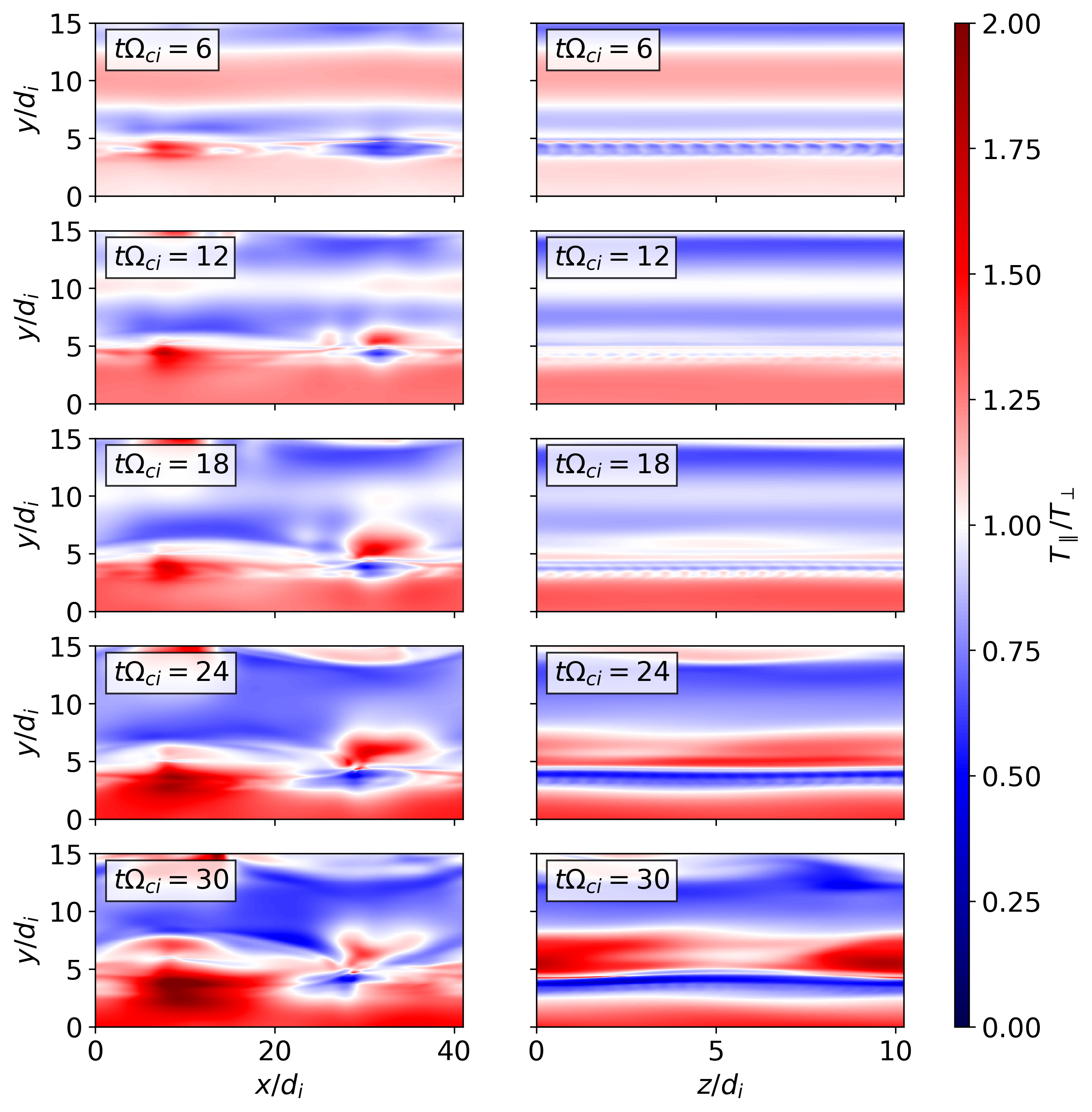}
    \caption{\label{fig:tparperpx} Temperature anisotropy in the reconnection plane (left), and along the X-line (right) at $z = L_z/2$ for the gradient closure. Gradient closure dynamics driven by the temperature gradient across the separatrices lead to anisotropies developing throughout both layers.}
\end{figure}

Fig.~\ref{fig:burch_temp} shows a high-temperature region along the separatrix and near the X-point for the gradient closure, similar to what is seen in the \citeA{le2017enhanced} kinetic results, though with a couple of distinct differences. Parallel temperature drops on the magnetosphere side of the X-point but is high on the magnetosheath side; however, the perpendicular temperature has the opposite behavior in both regions. The result is temperature anisotropy profiles from the gradient closure (Fig.~\ref{fig:tparperpx}) which differs from the expectation, as kinetic results and the MMS data both predict a temperature anisotropy that is high on the magnetosphere side. A similar discrepancy in the location of the temperature anisotropy was noted in the local closure results reported by \citeA{tenbarge2019extended}.\par
Also notable in the gradient closure results is the strong anisotropy throughout both the magnetosheath and magnetosphere layers. This anisotropy is easily accounted for in the physics of the gradient closure. The strongest initial temperature gradients lie in the $y$-direction, which drives heat flow in that direction. As the gradient closure takes gradients in different directions into account, this means greater flow in $y$ than in $x$, where the temperature is close to uniform. This extensive anisotropy is unlikely to be physical, both because diffusion across field lines should be small physically, and because any collisional effects if present would likely be sufficient to isotropize the small gradients developing in the magnetosheath and magnetosphere regions which cause this anisotropy. The result of the gradient closure, however, is heat flowing from the hot magnetosphere to the cold magnetosheath in the perpendicular direction, while the parallel direction remains high, leading to a high ratio of $T_{\parallel}/T_{\perp}$ in the magnetosphere and low $T_{\parallel}/T_{\perp}$ in the magnetosheath. Near the X-point, there is a reversal of this trend with $T_{\parallel}/T_{\perp}$ dropping very low on the magnetosphere side while being flipped on the magnetosheath side.\par

\begin{figure}
    \centering
    \includegraphics[width=0.8\linewidth]{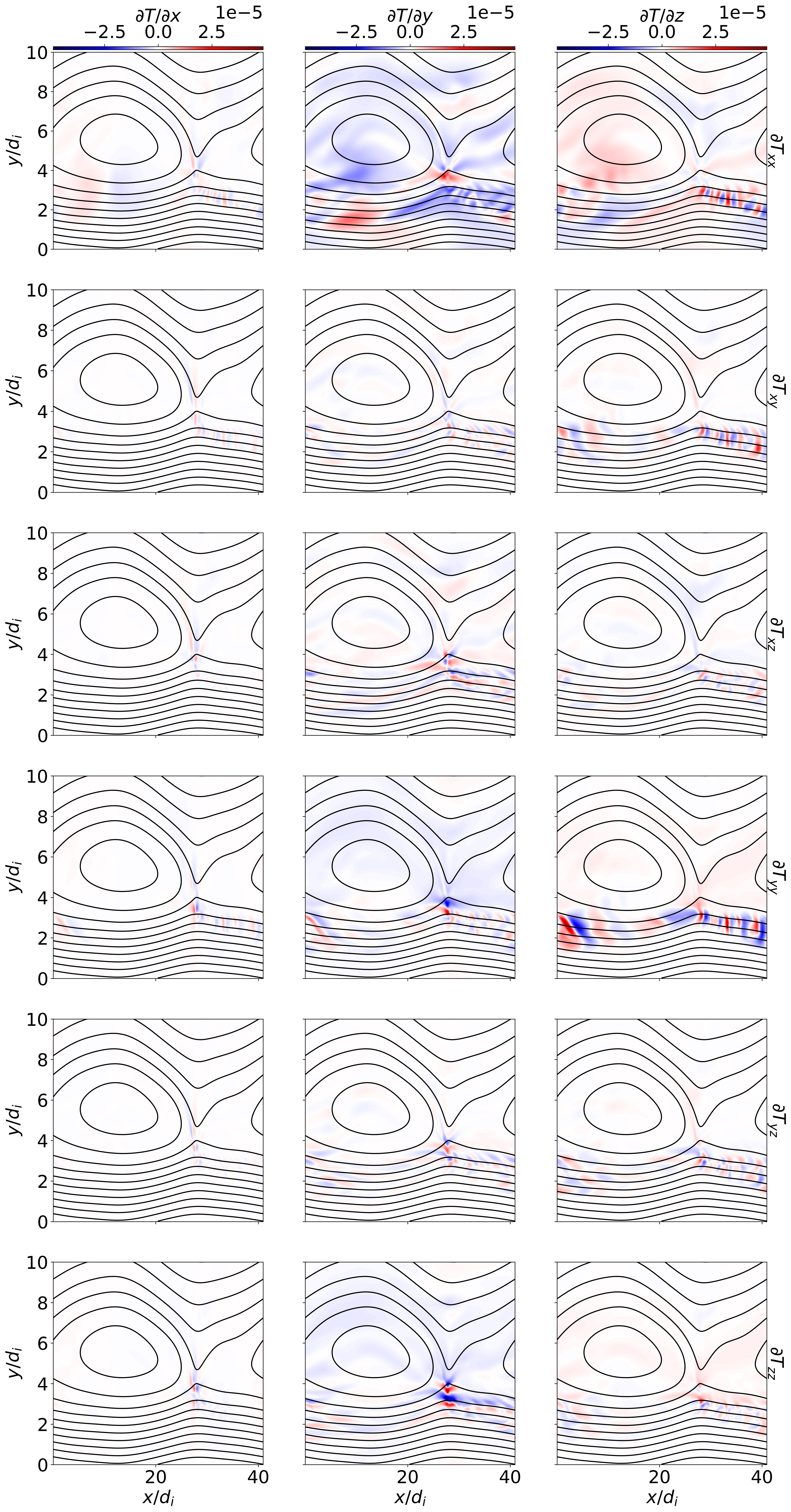}
    \caption{\label{fig:grad_update} Temperature gradients across the reconnection layer for the gradient closure at $t\Omega_{ci} = 30$. There is a gradient in the $y$-direction across the entire layer, which drives the extension of temperature anisotropy throughout the domain (see Fig.~\ref{fig:tparperpx}). The strongest gradients are in the $z$-direction at the separatrices.}
\end{figure}

Fig.~\ref{fig:grad_update} shows the temperature gradients driving the heat flux evolution by the gradient closure. Setting aside the $y$-gradient source of the extensive temperature anisotropy, which we have demonstrated to be unphysical, there is clear demonstration that it is typically gradients in the $z$-direction, driven by the transverse instabilities, that dominate temperature dynamics near the separatrices.\par

\section{Discussion}\label{sec:discussion}

Ten-moment models provide a far more complete system for simulating space plasmas than MHD, capturing many of the kinetic effects which are lost when neglecting electron inertial effects and the pressure tensor. The gradient closure gives a marked improvement on the previously employed local closure in its ability to properly capture transverse instabilities in the current sheet such as the initial LHDI and later kinking instabilities, and the model captures the anisotropic diffusion of the layer better than the local closure.\par
Use of the gradient closure incurs a significantly increased computational cost over the local closure. This cost is due to a combination of the gradient closure requiring all neighboring gradients to be evaluated and compared (through the limiter) at each vertex (see \ref{app:closure}), exacerbated by a significantly more restrictive time step than the local closure. Overall computational cost of the gradient closure simulation is around three times that of the local closure simulation for this parameter space.\par
Improvements on the current closure are still possible. The gradient closure, while capturing anisotropic diffusion due to varying directional temperature gradients, does not take into account the preferred direction along the magnetic field lines. The required modification to the closure to do this would be to solve an equation taking the form of the tensor equation
\begin{equation}
    \mathbf{Q} + \omega\mathbf{b}\times\mathbf{Q} = \mathbf{S},
\end{equation}
where the source term $S_{ijk} = -\frac{v_t}{|k_0|}\partial_{[i}T_{jk]}$ is the current formulation of the heat flux used by the gradient closure, and $\omega$ is a parameter proportional to the cyclotron frequency. This form would restrict the diffusion perpendicular to field lines with increasing $\omega$. Recent developments in ML-based closures for multifluid models \cite{donaghy2023,huang2025} may yield a more optimal closure, particularly when combined with this physics insight into the expected parallel versus perpendicular heat flux. However, it will be important that these modifications to the gradient-based closure still respect the second law of thermodynamics and that heat flows down temperature gradients. Thus, the combination of the magnetic field and any ML-based approach will need a precise generalization of the limiter deployed in this work to insure that each diagonal component of the pressure tensor stays positive-definite.\par
Two of the most notable divergences of the gradient closure from canonical results are the temperature anisotropy and agyrotropy of the pressure tensor, both of which extend across significantly greater portions of the layers than kinetic results indicate, rather than being confined near the X-point and along separatrices. While part of the purpose of the gradient closure was to capture the full range of anisotropy and agyrotropy in the pressure tensor that was being suppressed by the local closure, the extent to which both of these are present throughout the domain suggests \textit{some} collisional relaxation is probably necessary. There is significant potential for improvement through the simple addition of a small isotropizing term, of a lower magnitude than previous uses of the local closure, which could damp small deviations from isotropy and prevent them from perpetuating through the gradient closure without interfering with the better fidelity of the pressure tensor dynamics in the dissipation region. In other words, both closures together may provide further improved physics fidelity over the exclusive use of either.\par

\begin{figure}
    \includegraphics[width=1.0\linewidth]{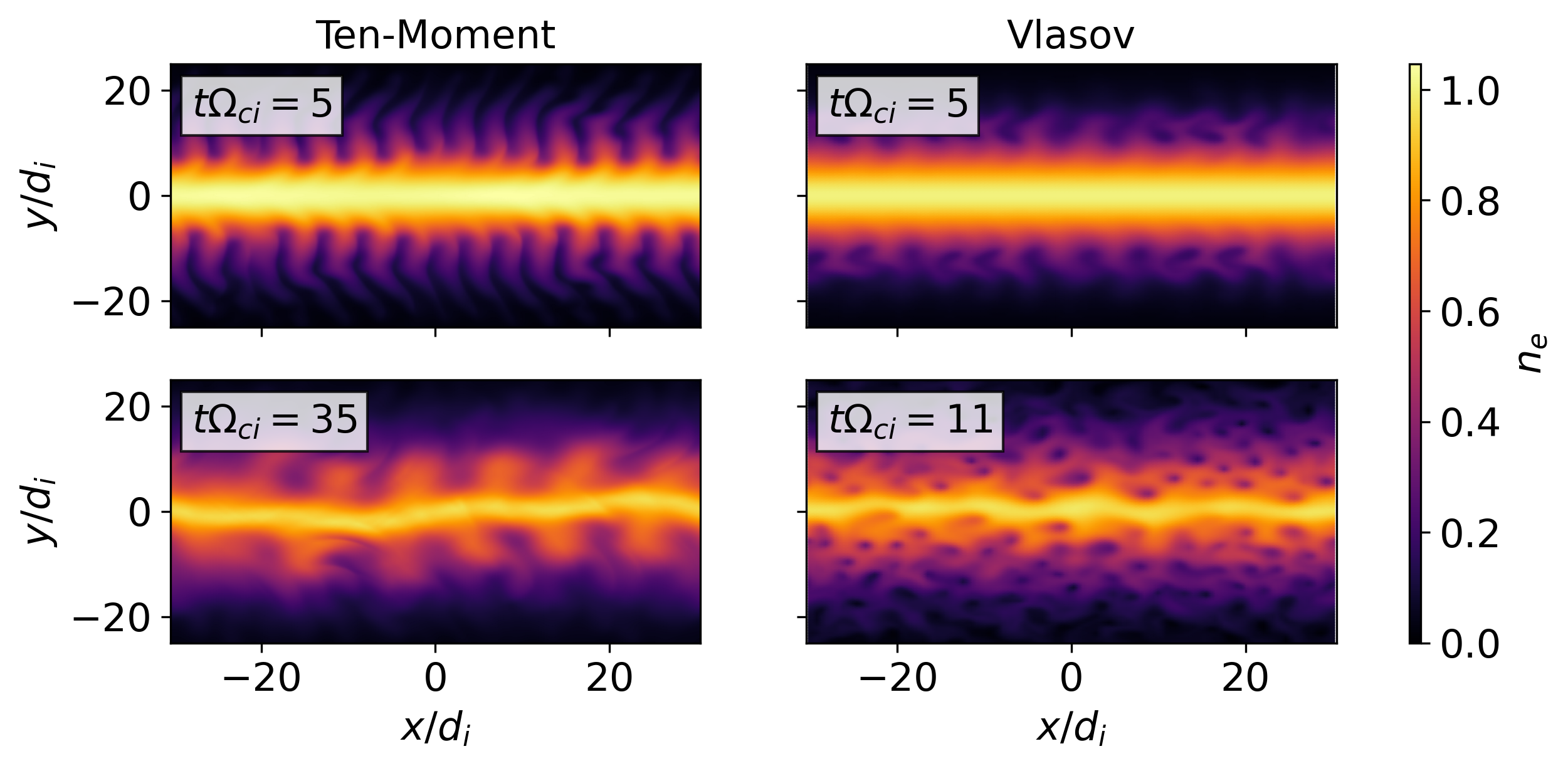}
    \caption{\label{fig:lhdi_dist} Comparison of LHDI in a simple 2D Harris sheet for ten-moment fluids (left) and Vlasov-Maxwell continuum-kinetics (right). The initial LHDI growth is similar for both fluid and kinetic results, while the drift-kink instability which grows during the nonlinear saturation grows earlier in the kinetic results and is accompanied by the growth of additional small-scale modes.}
\end{figure}

The linear growth of the LHDI is very well captured by the gradient closure, a noticeable improvement over the local closure, where the instability is not captured at all in this parameter regime for the chosen $k_0$. LHDI modes are excited in a current sheet when a cross-field current acts across a density gradient in the sheet \cite{daughton1999unstable}. The LHDI is often observed to trigger additional drift-kink modes \cite{yoon2002generalized,daughton2003electromagnetic} as seen in the 3D reconnection simulations. Past analysis of the multifluid model by \citeA{ng2019drift} using nonlocal closures demonstrated good capture of the LHDI and drift kink compared to kinetic results for reconnection regimes, while additional simulations in \citeA{ng2020improved} tested the gradient closure on simulations of LHDI with comparisons to kinetic results, but they did not examine the nonlinear saturation of the instability and subsequent drift-kink growth.\par
Due to the prominence of the secondary drift-kink instability in these simulations and the importance of gauging how well our choice of parameter $k_0$ succeeds in replicating the kinetic predictions, simple 2D Harris sheet cases with a comparable initialization to the Burch simulations were run in \texttt{Gkeyll} to compare the fluid model's ability to capture the kinking and equivalent continuum-kinetic Vlasov \cite{juno2018discontinuous,hakim2020alias} \texttt{Gkeyll} results. These results are shown in Fig.~\ref{fig:lhdi_dist}. Standard early growth of the LHDI occurs on comparable time scales in both, with a drift mode growing out of the nonlinear saturation later in time. Similar behavior is observed in both the kinetic and fluid results, though additional small-scale modes are captured by the kinetic simulation. Most notably, the time scale for the kinetic kink instability growing is significantly shorter in the Vlasov simulation, with it appearing within a few ion cyclotron periods of the LHDI onset. The fluid results take some tens of cyclotron periods to develop to the same level, similar to what was observed in the 3D reconnection simulations. This result indicates a potential expectation of more rapid and turbulent onset of secondary instabilities in kinetic reconnection results, but that the major features of the instability are well-captured regardless for this choice of parameter.\par

\section{Summary and Conclusions}\label{sec:conclusion}

We have presented results from a study of the \citeA{burch2016electron} 16 October 2015 MMS diffusion region crossing event using \texttt{Gkeyll}'s two-fluid, ten-moment model. The ten-moment model self-consistently captures the electron inertia and full pressure tensor dynamics. The primary closure employed for the system was the gradient-based model for the heat flux propoesed by \citeA{ng2020improved}, which was compared with a local relaxation closure employed by a previous study performed by \citeA{tenbarge2019extended}. The gradient closure was improved from prior implementations to use a symmetric scheme with a limiter ensuring that direction of heat flow is physically correct at the vertices.\par
The gradient closure was demonstrated to capture multiple features predicted by kinetic simulations which the local closure could not, most predominantly transverse LHDI and drift-kink instabilities that grow in the current sheet. However, temperature anisotropy and agyrotropy in the pressure tensor were found to be significantly more extensive than fully kinetic predictions, suggesting that an isotropizing term for the pressure tensor may be necessary.\par
Additional improvements to the closure are possible. In the collisionless limit, it is expected that the bulk of the temperature diffusion of the plasma will be in the direction parallel to magnetic fields, as the perpendicular flow will be restricted by the gyromotion of the particles. While the anisotropic diffusion with respect to magnetic field direction was captured better in the gradient closure than in the local closure, further improvement is possible through an additional term in the closure equation following Eq.~\ref{eq:ani_closure} to actively restrict cross-field flows. A synthesis of the local and gradient closure may also be necessary to damp out minor anistropy and agyrotropy in the pressure tensor, allowing upstream isotropization without preventing their development and propagation in the reconnection layer.\par
Future work will explore further increasing the fidelity of the closure, as well as application to global simulations of planetary magnetospheres. Past multifluid simulations of magnetospheres with the local closure have already demonstrated significant strides in making global modeling with improved electron dynamics affordable \cite{wang2018electron,dong2019global,jarmak2020quest}. Application and improvement of the gradient closure to these global simulations has the potential to further increase the physics fidelity of the multifluid model, maximizing their ability to capture kinetic physics without prohibitive computational cost.

\appendix
\section{Implementation of the Gradient Closure}\label{app:closure}

\begin{figure}
	\centering
	\begin{tikzpicture}
		\matrix[matrix of nodes, nodes={minimum size=2.25cm, draw, anchor=center}, row sep=-\pgflinewidth, column sep=-\pgflinewidth](mygrid){%
		$T^{m-1,n+1}_{ij}$ & $T^{m,n+1}_{ij}$ & $T^{m+1,n+1}_{ij}$ \\
		$T^{m-1,n}_{ij}$ & $T^{m,n}_{ij}$ & $T^{m+1,n}_{ij}$ \\
		$T^{m-1,n-1}_{ij}$ & $T^{m,n-1}_{ij}$ & $T^{m+1,n-1}_{ij}$ \\
		};
	\end{tikzpicture}
	\caption{\label{fig:comp_grid} 2D 3x3 computational grid for demonstrating the implementation of the gradient closure.}
\end{figure}

Using Fig.~\ref{fig:comp_grid} for reference, we define a 3x3 2D computational grid with components of the temperature tensor $T_{ij} = P_{ij}/n$ specified at the cell centers. The closure equation for the heat flux tensor (Eq.~\ref{eq:grad_closure}) expands to
\begin{equation}
	Q_{ijk} = -\frac{v_t}{|k_0|}\partial_{[k}T_{ij]} = -\frac{1}{3}\frac{v_t}{k_0}\bigg(\frac{\partial T_{ij}}{\partial x_k} + \frac{\partial T_{jk}}{\partial x_i} + \frac{\partial T_{ki}}{\partial x_j}\bigg).\label{eq:symm_closure}
\end{equation}
Taking the cell at indices $m$ and $n$, we can calculate the four neighboring gradients around the cell corner at $n+1/2$, $m+1/2$ to be
\begin{equation}
\Delta_x T_{ij}^{m+1/2,n} = \frac{T^{m+1,n}_{ij} - T^{m,n}_{ij}}{\Delta x},
\end{equation}
\begin{equation}
\Delta_x T_{ij}^{m+1/2,n+1} = \frac{T^{m+1,n+1}_{ij} - T^{m,n+1}_{ij}}{\Delta x},
\end{equation}
\begin{equation}
\Delta_y T_{ij}^{m,n+1/2} = \frac{T^{m,n+1}_{ij} - T^{m,n}_{ij}}{\Delta y},
\end{equation}
\begin{equation}
\Delta_y T_{ij}^{m+1,n+1/2} = \frac{T^{m+1,n+1}_{ij} - T^{m+1,n}_{ij}}{\Delta y}.
\end{equation}
As detailed by \citeA{sharma2007preserving}, a direct average of the upper and lower gradients to approximate the cell corner gradient can cause heat flow in the wrong direction in neighboring cells. To prevent this, we adopt the limiter employed by \citeA{sharma2007preserving} for symmetric methods,
\begin{equation}
	L(a,b) = \begin{cases}
				(a + b)/2 & \text{if }\min{(\alpha a, a/\alpha)} < (a + b)/2 < \max{(\alpha a, a/\alpha)},\\
	             \min{(\alpha a, a/\alpha)} & \text{if }(a + b)/2 \leq \min{(\alpha a, a/\alpha)},\\
	             \max{(\alpha a, a/\alpha)} & \text{if }(a + b)/2 \geq \max{(\alpha a, a/\alpha)},
	         \end{cases}
\end{equation}
using the parameter $\alpha=0.75$. The upper and lower cell corner gradients may then be separated by the limiter, becoming
\begin{equation}
\frac{\partial T_{ij}}{\partial x}\bigg\rvert_{m+1/2,n+1/2}^+ = L(\Delta_x T_{ij}^{m+1/2,n+1}, \Delta_x T_{ij}^{m+1/2,n}),
\end{equation}
\begin{equation}
\frac{\partial T_{ij}}{\partial x}\bigg\rvert_{m+1/2,n+1/2}^- = L(\Delta_x T_{ij}^{m+1/2,n}, \Delta_x T_{ij}^{m+1/2,n+1}),
\end{equation}
\begin{equation}
\frac{\partial T_{ij}}{\partial y}\bigg\rvert_{m+1/2,n+1/2}^+ = L(\Delta_y T_{ij}^{m+1,n+1/2}, \Delta_y T_{ij}^{m,n+1/2}),
\end{equation}
\begin{equation}
\frac{\partial T_{ij}}{\partial y}\bigg\rvert_{m+1/2,n+1/2}^- = L(\Delta_y T_{ij}^{m+1,n+1/2}, \Delta_y T_{ij}^{m,n+1/2}).
\end{equation}
Eq.~\ref{eq:symm_closure} is then expanded into the computational domain to become
\begin{equation}
	Q_{ijk}^{m+1/2,n+1/2} = -\frac{1}{3}\frac{\tilde{v}_t}{k_0}\bigg(\frac{\partial T_{ij}}{\partial x_k}\bigg\rvert_{m+1/2,n+1/2}^{\pm} + \frac{\partial T_{jk}}{\partial x_i}\bigg\rvert_{m+1/2,n+1/2}^{\pm} + \frac{\partial T_{ki}}{\partial x_j}\bigg\rvert_{m+1/2,n+1/2}^{\pm}\bigg),
\end{equation}
where $\tilde{v}_t$ represents the harmonic average of the thermal speed at the corner, and the sign of $\pm$ depends on whether the cell being updated is the upper or lower cell in the gradient direction. The final contribution of the closure to the source term in the cell is then taken be
\begin{equation}
	\frac{\partial Q^{m,n}_{ijk}}{\partial x_k} = \frac{Q_{ijk}^{m+1/2,n+1/2} \pm Q_{ijk}^{m+1/2,n-1/2} \mp Q_{ijk}^{m-1/2,n+1/2} - Q_{ijk}^{m-1/2,n-1/2}}{4\Delta x_k},
\end{equation}
an averaging of the gradients across the cell. The 3D implementation follows directly from a generalization of these equations into an additional dimension, bringing the number of neighboring gradients at the vertex from four to eight.

\section*{Open Research Section}

All the simulation results presented in this paper were produced by and are reproducible using the open-source \texttt{Gkeyll} software. Information for obtaining, installing, and running \texttt{Gkeyll} may be found on the documentation site (\url{https://gkeyll.readthedocs.io}). The input files for the simulations used to produce the results in this paper may be acquired from the repository at \url{https://github.com/ammarhakim/gkyl-paper-inp/tree/master/2025_JGR_Burch}.

\acknowledgments

The work presented here was supported by the NSF Collaborative Research: Frameworks: A Software Ecosystem for Plasma Science and Space Weather Applications project under Award Number 2209471.\par
The simulations presented in this article were performed on computational resources managed and supported by Princeton Research Computing, a consortium of groups including the Princeton Institute for Computational Science and Engineering (PICSciE) and the Office of Information Technology’s High Performance Computing Center and Visualization Laboratory at Princeton University.

%%%%%%%%%%%%%%%%%%%%%%%%%%%%%%%%%%%%%%%%%%%%%%%
% REFERENCES and BIBLIOGRAPHY
%
\bibliography{references}
%
%%%%%%%%%%%%%%%%%%%%%%%%%%%%%%%%%%%%%%%%%%%%%%%

\end{document}